\documentclass[a4paper,11pt]{article}
\pdfoutput=1
\usepackage{pos}

\title{Isoscalar electromagnetic form factors of the nucleon in \texorpdfstring{$N_\mathrm{f} = 2 + 1$}{Nf=2+1} lattice QCD}

\author[a,b]{Dalibor Djukanovic}
\author[c]{Georg von Hippel}
\author[a,b,c]{Harvey B. Meyer}
\author[c]{Konstantin Ottnad}
\author*[c]{Miguel Salg}
\author[c]{Jonas Wilhelm}
\author[a,b,c]{Hartmut Wittig}

\affiliation[a]{Helmholtz Institute Mainz, Staudingerweg 18, 55128 Mainz, Germany}
\affiliation[b]{GSI Helmholtzzentrum für Schwerionenforschung, 64291 Darmstadt, Germany}
\affiliation[c]{\texorpdfstring{PRISMA${}^+$}{PRISMA+} Cluster of Excellence and Institute for Nuclear Physics, Johannes Gutenberg University of Mainz, Johann-Joachim-Becher-Weg 45, 55128 Mainz, Germany}

\emailAdd{d.djukanovic@him.uni-mainz.de}
\emailAdd{hippel@uni-mainz.de}
\emailAdd{meyerh@uni-mainz.de}
\emailAdd{kottnad@uni-mainz.de}
\emailAdd{msalg@uni-mainz.de}
\emailAdd{jwilhe@uni-mainz.de}
\emailAdd{hartmut.wittig@uni-mainz.de}

\abstract{
We present results for the isoscalar electromagnetic form factors of the nucleon computed on the Coordinated Lattice Simulations (CLS) ensembles with $N_\mathrm{f} = 2 + 1$ flavors of $\mathcal{O}(a)$-improved Wilson fermions and an $\mathcal{O}(a)$-improved conserved vector current.
In order to estimate the excited-state contamination, we employ several source-sink separations and apply the summation method.
For the computation of the quark-disconnected diagrams, a stochastic estimation based on the one-end trick is performed, in combination with a frequency-splitting technique and the hopping parameter expansion.
By these means, we obtain a clear signal for the form factors including the quark-disconnected contributions, which have a statistically significant effect on our results.
\vspace{0.5cm}
\begin{flushright}
    MITP-21-049
\end{flushright}
}

\FullConference{%
 The 38th International Symposium on Lattice Field Theory, LATTICE2021\\
  26th-30th July, 2021\\
  Zoom/Gather@Massachusetts Institute of Technology
}

\makeatletter
\hypersetup{pdftitle=\@title, pdfauthor=\printHeadAuthors}
\makeatother
\usepackage[USenglish]{babel}
\usepackage[group-digits=integer, exponent-product= \cdot]{siunitx}
\usepackage{csquotes}
\usepackage[english]{cleveref}
\usepackage[all]{foreign}
\usepackage{booktabs}
\usepackage{tikz}
\usepackage{braket}
\usepackage{subcaption}
\DeclareMathOperator{\tr}{tr}

\begin{document}
\maketitle

\section{Introduction}
The internal structure of the nucleon is still an open research field in subatomic physics.
In particular, there is a discrepancy between different measurements of the electric charge radius of the proton:
The value obtained from $ep$-scattering \cite{Bernauer2014}, while in good agreement with hydrogen spectroscopy \cite{Mohr2012}, is incompatible with the most accurate determination from the spectroscopy of muonic hydrogen \cite{Antognini2013}.
Hence, the electromagnetic form factors of the proton and neutron, from which the radius is extracted in the context of scattering experiments, are of a lasting and high interest in the community.

For our theoretical calculations, we split the form factors up into an isovector and an isoscalar part.
Whereas the former only contains quark-connected contributions, in the latter also quark-disconnected diagrams appear.
A full prediction of the proton and neutron form factors from first principles therefore necessitates an explicit treatment of isoscalar quantities on the lattice, including the disconnected contributions.
Following our recent publication of the isovector electromagnetic form factors \cite{Djukanovic2021}, we present here the current state of our lattice QCD-based determination of the isoscalar electromagnetic form factors of the nucleon from the $N_\mathrm{f} = 2 + 1$ CLS ensembles \cite{Bruno2015}.
Our preliminary results point towards a small value of the electric charge radius of the proton, consistent with the findings in \cite{Djukanovic2021}, although it is still too early to draw any definite conclusions.

These proceedings are organized as follows:
\Cref{sec:setup} deals with the ensembles and operators used in our simulations, while \cref{sec:methods} is dedicated to the methods employed to extract the form factors and charge radii from our lattice data.
In \cref{sec:results} we present our preliminary results reflecting the current state of the analysis.
\Cref{sec:conclusions} finally draws some conclusions and gives an outlook to further planned work on this project.

\section{Lattice setup}
\label{sec:setup}
We use the CLS ensembles \cite{Bruno2015} which have been generated with $2 + 1$ flavors of non-perturbatively $\mathcal{O}(a)$-improved Wilson fermions \cite{Sheikholeslami1985,Bulava2013} and a tree-level improved Lüscher-Weisz gauge action \cite{Luescher1985}.
Only ensembles following the $\tr M_q = 2m_l + m_s = \text{const.}$ trajectory are employed.
In order to prevent topological freezing, the fields obey open boundary conditions in time, with the exception of the near-physical ensemble E250, which uses periodic boundary conditions in time.
\Cref{tab:ensembles} displays the current set of ensembles:
We have already analyzed data at three lattice spacings covering the range from \SI{0.050}{fm} to \SI{0.086}{fm}, three different spatial extents at the intermediate value of $a$, and several different pion masses, including one slightly below the physical value (E250).

\begin{table}[ht]
    \centering
    \begin{tabular}{l@{\hspace{5mm}}cccccccc}
        \toprule
        ID   & $\beta$ & $a$ [fm]      & $N_\tau$ & $N_s$ & $M_\pi$ [MeV] & $N_\mathrm{cfg}^\mathrm{conn}$ & $N_\mathrm{cfg}^\mathrm{disc}$ \\ \midrule
        C101 & 3.40    & 0.08636(106)  & 96       & 48    & 223(3)        & 2000                           & 1000                           \\[\defaultaddspace]
        N200 & 3.55    & 0.06426(76)   & 128      & 48    & 283(3)        & 1712                           & 1712                           \\
        D200 & 3.55    & 0.06426(76)   & 128      & 64    & 203(3)        & 1108                           & 554                            \\
        E250 & 3.55    & 0.06426(76)   & 192      & 96    & 130(1)        & 250                            & 250                            \\[\defaultaddspace]
        J303 & 3.70    & 0.04981(57)   & 192      & 64    & 262(3)        & 1073                           & 1073                           \\ \bottomrule
    \end{tabular}
    \caption{Overview of the currently analyzed ensembles. The quoted errors on the pion masses include the error from the scale setting \cite{Bruno2017}.}
    \label{tab:ensembles}
\end{table}

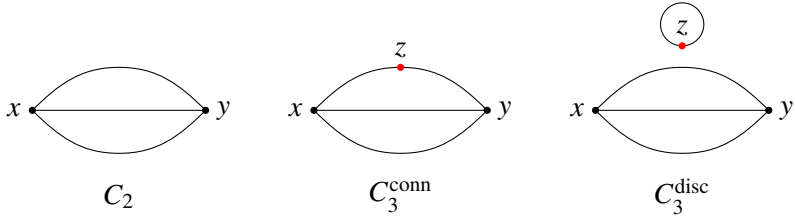
\begin{figure}[ht]
    \begin{center}
        \begin{tikzpicture}[scale=0.57]
	\node[left] at (-2,0) {$x$};
	\node[right] at (2,0) {$y$};
	
	\filldraw (-2,0) circle (0.07) -- (2,0) circle (0.07);
	\draw (-2,0) to [out=45, in=180] (0,1) to [out=0, in=135] (2,0);
	\draw (-2,0) to [out=315, in=180] (0,-1) to [out=0, in=225] (2,0);
	
	\node[left] at (4.5,0) {$x$};
	\node[right] at (8.5,0) {$y$};
	\node[above] at (6.5,1) {$z$};
	
	\filldraw (4.5,0) circle (0.07) -- (8.5,0) circle (0.07);
	\draw (4.5,0) to [out=45, in=180] (6.5,1) to [out=0, in=135] (8.5,0);
	\filldraw[color=red] (6.5,1) circle (0.07);
	\draw (4.5,0) to [out=315, in=180] (6.5,-1) to [out=0, in=225] (8.5,0);
	
	\node[left] at (11,0) {$x$};
	\node[right] at (15,0) {$y$};
	\node[above] at (13,1.5) {$z$};
	
	\filldraw (11,0) circle (0.07) -- (15,0) circle (0.07);
	\draw (11,0) to [out=45, in=180] (13,1) to [out=0, in=135] (15,0);
	\draw (13,2) circle (0.5);
	\filldraw[color=red] (13,1.5) circle (0.07);
	\draw (11,0) to [out=315, in=180] (13,-1) to [out=0, in=225] (15,0);
	
	\node at (0,-2) {$C_2$};
	\node at (6.5,-2) {$C_3^\mathrm{conn}$};
	\node at (13,-2) {$C_3^\mathrm{disc}$};
\end{tikzpicture}
    \end{center}
    \caption{Diagrammatic representation of the two- and three-point functions of the nucleon. Only quark lines are shown, while all gluon lines are suppressed. The red dots in the three-point functions represent the operator insertion.}
    \label{fig:C2_3}
\end{figure}

Concretely, we measure the two- and three-point functions of the nucleon, which are depicted diagrammatically in \cref{fig:C2_3},
\begin{align}
    \label{eq:C2}
    C_2(\mathbf{p'}; y_0, x_0) &= \sum_{\mathbf{y}} e^{-i\mathbf{p'y}} \Gamma_{\beta\alpha} \Braket{0 | N_\alpha(\mathbf{y}, y_0) \bar{N}_\beta(\mathbf{0}, x_0) | 0} , \\
    \label{eq:C3_connected}
    C_{3, O}(\mathbf{p'}, \mathbf{q}; y_0, z_0, x_0) &= \sum_{\mathbf{y}, \mathbf{z}} e^{i \mathbf{qz}} e^{-i\mathbf{p'y}} \Gamma_{\beta\alpha} \Braket{0 | N_\alpha(\mathbf{y}, y_0) O(\mathbf{z}, z_0) \bar{N}_\beta(\mathbf{0}, x_0) | 0} , \\
    \label{eq:C3_disconnected}
    C_{3, O}^\mathrm{disc}(\mathbf{p'}, \mathbf{q}; y_0, z_0, x_0) &= \left\langle L^{O, \mathrm{disc}}(\mathbf{q}; z_0) C_2(\mathbf{p'}; y_0, x_0) \right\rangle , \\
    \label{eq:loops}
    L^{O, \mathrm{disc}}(\mathbf{q}; z_0) &= -\sum_{\mathbf{z}} e^{i \mathbf{q} \mathbf{z}} \tr[S(z, z) \Gamma] .
\end{align}
Here, the same projection matrix $\Gamma = \frac{1}{2} (1 + \gamma_0) (1 + i \gamma_5 \gamma_3)$ is employed for both the two- and three-point functions, ensuring that the two of them are fully correlated.
For the connected part of the three-point functions, the nucleon at the sink is at rest, \ie for a momentum transfer \textbf{q} the initial and final states have momenta $\mathbf{p'} = \mathbf{0}$ and $\mathbf{p} = -\mathbf{q}$, respectively.
The disconnected part of the three-point functions is constructed from the quark loops and the two-point functions according to \cref{eq:C3_disconnected}.
The all-to-all propagator $S(z, z)$ appearing in the quark loops \cref{eq:loops} is computed via a stochastic estimation using a frequency-splitting technique \cite{Giusti2019}.
To that end, we employ a hopping-parameter expansion for one heavy quark flavor and subsequently apply the one-end trick for the remaining flavors.
Furthermore, we average over the forward and backward propagating nucleon and over several different sink momenta ($\mathbf{n}_{p'}^2 \leq 2$) for the disconnected contribution.

\begin{table}[ht]
    \centering
    \begin{tabular}{l@{\hspace{5mm}}cc@{\hspace{5mm}}cc@{\hspace{5mm}}c}
        \toprule
        ID   & $N_\mathrm{src, HP}^\mathrm{conn}$ & $N_\mathrm{src, LP}^\mathrm{conn}$ & $N_\mathrm{src, HP}^\mathrm{disc}$ & $N_\mathrm{src, LP}^\mathrm{disc}$ & $t_\mathrm{sep}/a$     \\ \midrule
        C101 & 1                                  & 32                                 & 6                                  & 192                                & 12, 14, 16             \\[\defaultaddspace]
        N200 & 1                                  & 12                                 & 1                                  & 12                                 & 16, 18, 20, 22         \\
        D200 & 1                                  & 32                                 & 7                                  & 224                                & 16, 18, 20, 22         \\
        E250 & 1, 1, 2, 4, 8                      & 16, 32, 64, 128, 256               & 8                                  & 256                                & 14, 16, 18, 20, 22     \\[\defaultaddspace]
        J303 & 1                                  & 16                                 & 1                                  & 16                                 & 20, 22, 24, 26         \\ \bottomrule
    \end{tabular}
    \caption{Number of high-precision (HP) and low-precision (LP) solves used for the computation of the connected and the disconnected contributions, respectively. For E250, the comma-separated numbers refer to the corresponding source-sink separations $t_\mathrm{sep}$.}
    \label{tab:sources}
\end{table}

To reduce the cost of the inversions, we apply the truncated-solver method \cite{Bali2010,Blum2013}.
Details on our setup of sources are collected in \cref{tab:sources}, alongside the available source-sink separations.
We use additional measurements of the two-point function for the disconnected contributions on the ensembles C101 and D200.
At the physical pion mass (E250), we employ iterative statistics for the different source-sink separations.
This means that with each step in $t_\mathrm{sep}$, the statistics for the connected part is doubled.
For the disconnected part, the highest statistics at our disposal is always utilized, in order to get the best signal.

As in \cite{Djukanovic2021}, we use a symmetrized conserved vector current, so that no renormalization is required.
The $\mathcal{O}(a)$-improvement is performed with the improvement coefficients computed in \cite{Gerardin2019a}.
The remaining technical aspects of our setup are identical to our previous papers \cite{Harris2019,Djukanovic2021}, to which we refer the interested reader.

\section{Extraction of the form factors and radii}
\label{sec:methods}
The present section is concerned with the methods used to extract the form factors and radii from our lattice data.
We proceed in three essential steps, which are presented in the following.

Starting from the two- and three-point functions \cref{eq:C2,eq:C3_connected,eq:C3_disconnected}, we calculate the ratios \cite{Korzec2009}
\begin{align}
    R_O(\mathbf{p'}, \mathbf{q}; t_\mathrm{sep}, t) = \frac{C_{3, O}(\mathbf{p'}, \mathbf{q}; t_\mathrm{sep}, t)}{C_2(\mathbf{p'}; t_\mathrm{sep})} \sqrt{\frac{C_2(\mathbf{p'}-\mathbf{q}; t_\mathrm{sep} - t) C_2(\mathbf{p'}; t) C_2(\mathbf{p'}; t_\mathrm{sep})}{C_2(\mathbf{p'}; t_\mathrm{sep} - t) C_2(\mathbf{p'}-\mathbf{q}; t) C_2(\mathbf{p'}-\mathbf{q}; t_\mathrm{sep})}} ,
    \label{eq:ratio}
\end{align}
where the source-sink separation is given by $t_\mathrm{sep} = y_0 - x_0$, and $t = z_0 - x_0$ denotes the temporal distance of the operator insertion from the source.
The two-point functions are averaged over equivalent momentum classes before plugging them into \cref{eq:ratio}.
At zero sink momentum (which is always the case for the connected contributions), the effective form factors can be calculated from the ratios \cref{eq:ratio} by simple analytical expressions \cite{Wilhelm2019,Djukanovic2021}.
If finite sink momenta are included, the effective form factors are extracted by solving a (generally overdetermined) system of equations following Refs.\@ \cite{Djukanovic2019,Wilhelm2019}.
Afterwards, the isoscalar (octet) combination $u+d-2s$ is put together from the connected and disconnected pieces as
\begin{equation}
    G_{E, M}^{\mathrm{eff},u+d-2s} = G_{E, M}^{\mathrm{eff,conn},u+d} + 2G_{E, M}^{\mathrm{eff,disc},l-s} .
    \label{eq:eff_ff_u+d-2s}
\end{equation}
Note that the disconnected part only requires the difference $l - s$ between the light and strange contributions, in which correlated noise cancels and which can be computed very effectively by the one-end trick.

In general, baryonic correlation functions suffer from a strong signal-to-noise problem at large Euclidean time separations \cite{Lepage1989}.
This necessitates an explicit treatment of the excited-state systematics in order to extract the ground-state form factors from the effective ones computed at the typically accessible source-sink separations.
In this work, we employ the \enquote{plain} (one-state) summation method \cite{Capitani2015,Djukanovic2021}.
We have checked explicitly that for the source-sink separations from \cref{tab:sources} the data do not deviate significantly from the corresponding linear behavior in $t_\mathrm{sep}$.

As a simple model for the $Q^2$-dependence of the form factors, we fit our data to a dipole \ansatz with a cut in $Q^2$ at \SI{0.6}{GeV^2}, so that at least four data points enter each fit.
$G_E(0)$ is fixed by fitting the normalized ratio $G_E(Q^2) / G_E(0)$.
Subsequently, the electric and magnetic charge radii and the magnetic moment \cite{Capitani2015} are extracted from the respective dipole forms.

\section{Preliminary results}
\label{sec:results}
In the following, we present some illustrative results obtained so far with the procedures explained in \cref{sec:setup,sec:methods}.

The isoscalar effective form factors as a function of the operator insertion time have been compared between two different setups: once without employing any additional measurements, \ie using the same statistics of the two-point functions for both the connected and the disconnected contributions, and once with the additional measurements for the disconnected part as detailed in \cref{tab:sources}.
On C101 and D200, the latter leads to a drastically improved signal for both the electric and the magnetic form factor.
We conclude that not only the statistics of the quark loops is relevant for the final quality of the data -- this has already been driven down to gauge noise -- but also the statistics of the two-point functions is crucial.
Note however that on some ensembles (like J303) we already achieve a decent signal without any additional measurements of the two-point functions.

\begin{figure}[ht]
    \begin{center}
        \begin{subfigure}{0.49\textwidth}
            \begin{center}
                \includegraphics[width=0.95\textwidth]{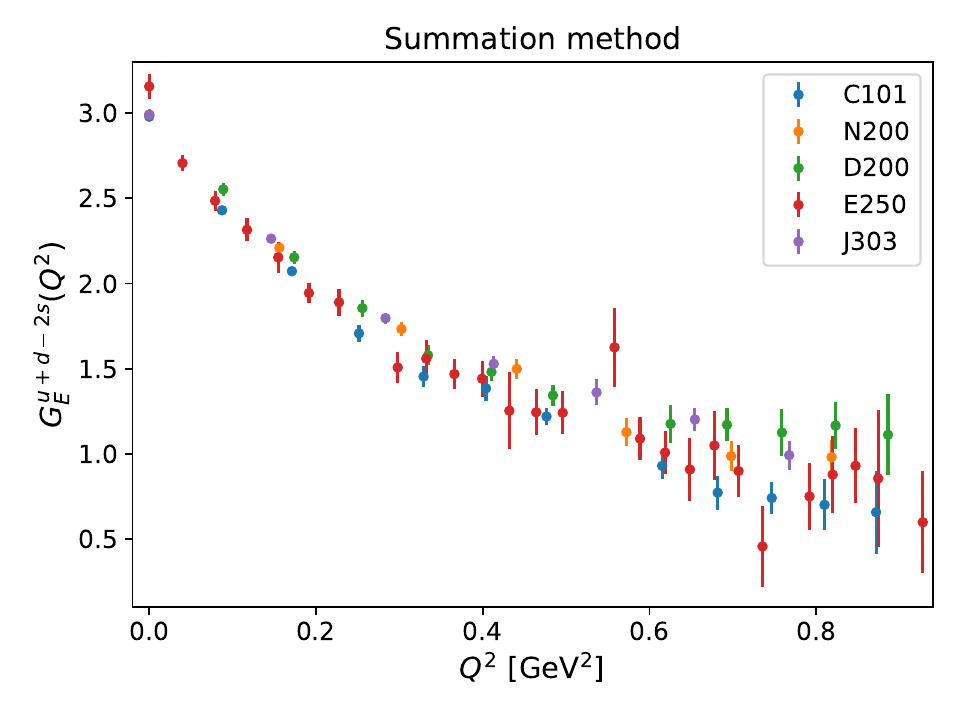}
            \end{center}
        \end{subfigure}
        \begin{subfigure}{0.49\textwidth}
            \begin{center}
                \includegraphics[width=0.95\textwidth]{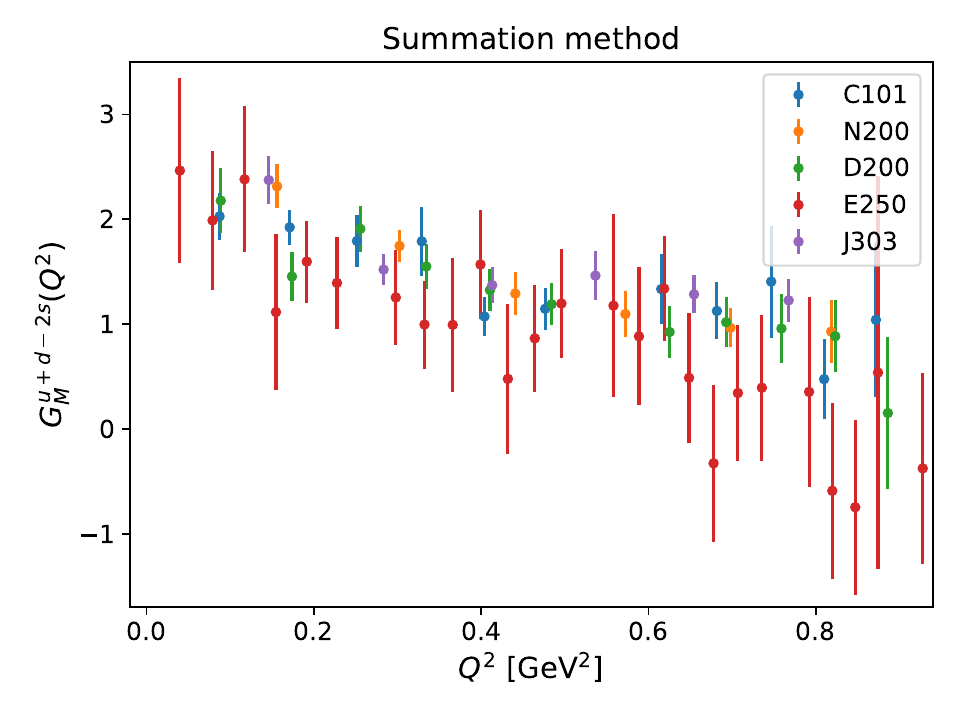}
            \end{center}
        \end{subfigure}
    \end{center}
    \caption{Isoscalar form factors as a function of $Q^2$: on the left the electric, on the right the magnetic form factor}
    \label{fig:ge_gm_isoscalar_combined}
\end{figure}

\begin{figure}[ht]
    \begin{center}
        \begin{subfigure}{0.49\textwidth}
            \begin{center}
                \includegraphics[width=0.91\textwidth]{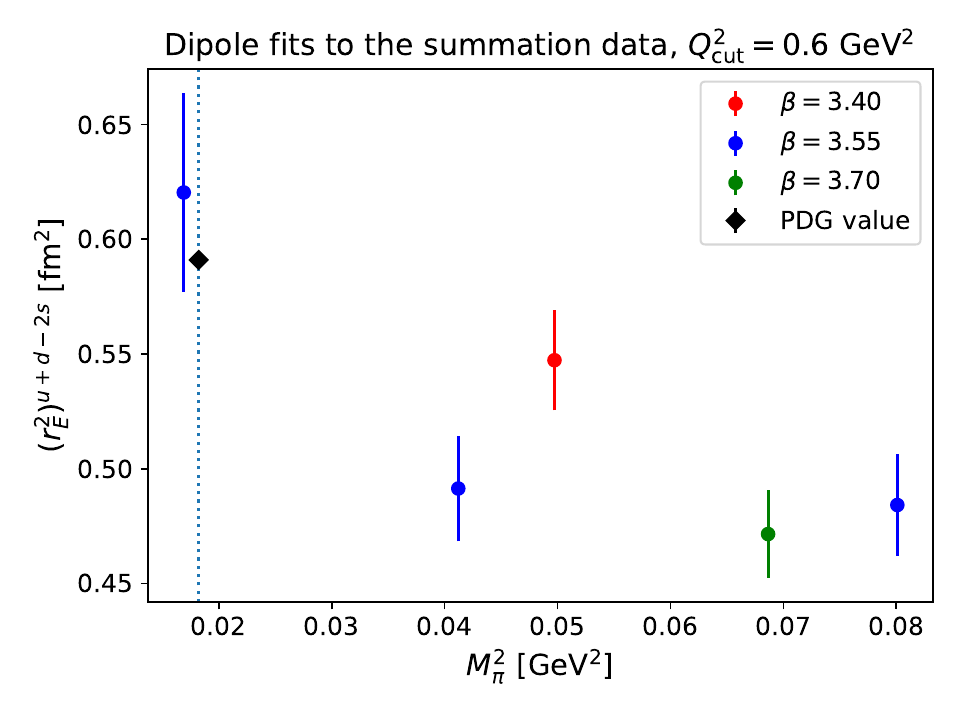}
            \end{center}
        \end{subfigure}
        \begin{subfigure}{0.49\textwidth}
            \begin{center}
                \includegraphics[width=0.91\textwidth]{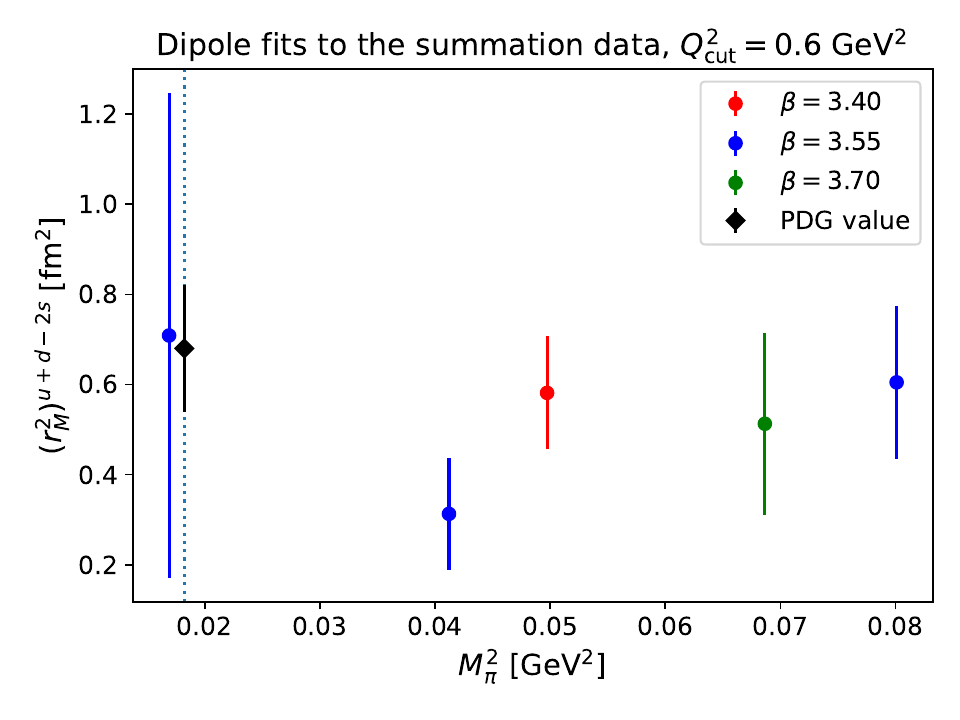}
            \end{center}
        \end{subfigure}
        \begin{subfigure}{0.49\textwidth}
            \begin{center}
                \includegraphics[width=0.91\textwidth]{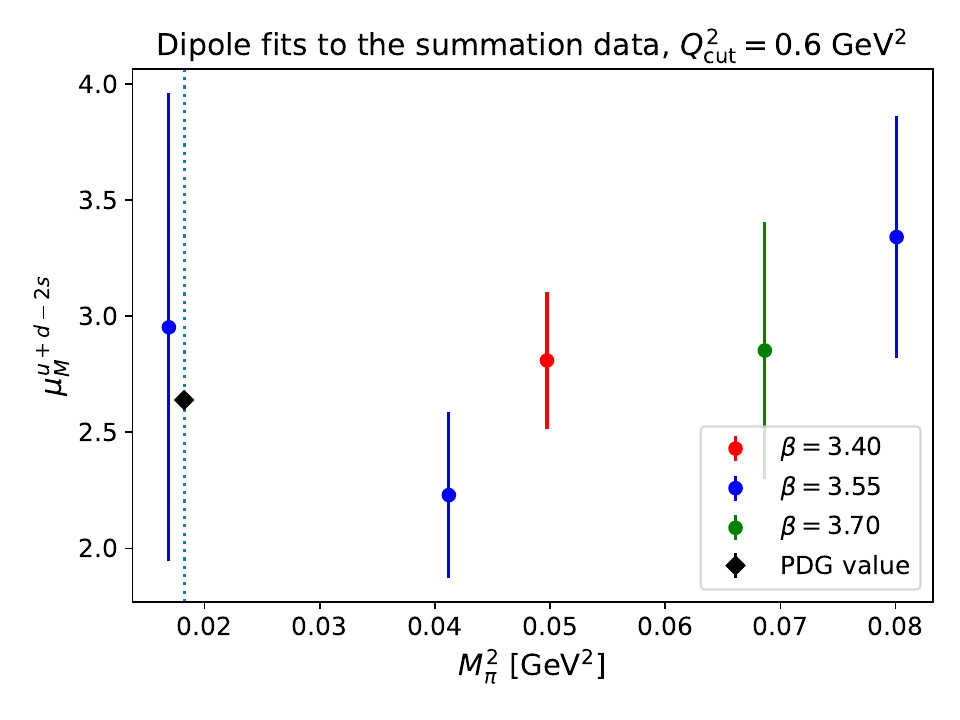}
            \end{center}
        \end{subfigure}
    \end{center}
    \caption{Isoscalar electric and magnetic charge radii and magnetic moment as a function of the pion mass squared. The experimental values have been calculated from the 2021 update of \cite{Zyla2020}.}
    \label{fig:isoscalar_radii}
\end{figure}

The $Q^2$-dependence of the ground-state isoscalar form factors as obtained from the summation method is displayed in \cref{fig:ge_gm_isoscalar_combined}, where we show combined plots for all five ensembles from \cref{tab:ensembles}.
Comparing with the pion masses given there, one can already see a chiral dependence of the form factors in these plots.
To judge this more accurately, it is nevertheless advantageous to look at the charge radii and the magnetic moment as a function of $M_\pi^2$.
Such plots can be found in \cref{fig:isoscalar_radii}, where the data points have been produced by dipole fits with a cut in $Q^2$ at \SI{0.6}{GeV^2}.
For all three quantities, a slight chiral trend is visible.
In the case of the electric radius, this also points approximately in the direction of our near-physical ensemble E250, which agrees within errors with the experimental value \cite{Zyla2020}.
For the magnetic quantities, however, the errors on E250 are still so large that this point cannot confirm any tendency yet.

\begin{figure}[ht]
    \begin{center}
        \begin{subfigure}{0.49\textwidth}
            \begin{center}
                \includegraphics[width=0.95\textwidth]{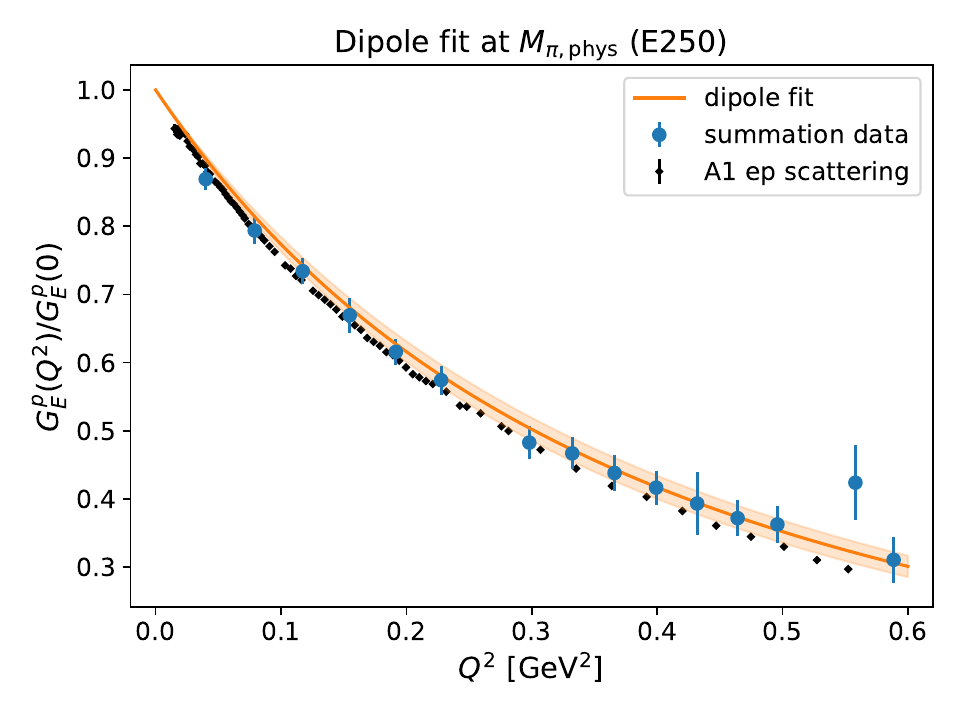}
            \end{center}
        \end{subfigure}
        \begin{subfigure}{0.49\textwidth}
            \begin{center}
                \includegraphics[width=0.95\textwidth]{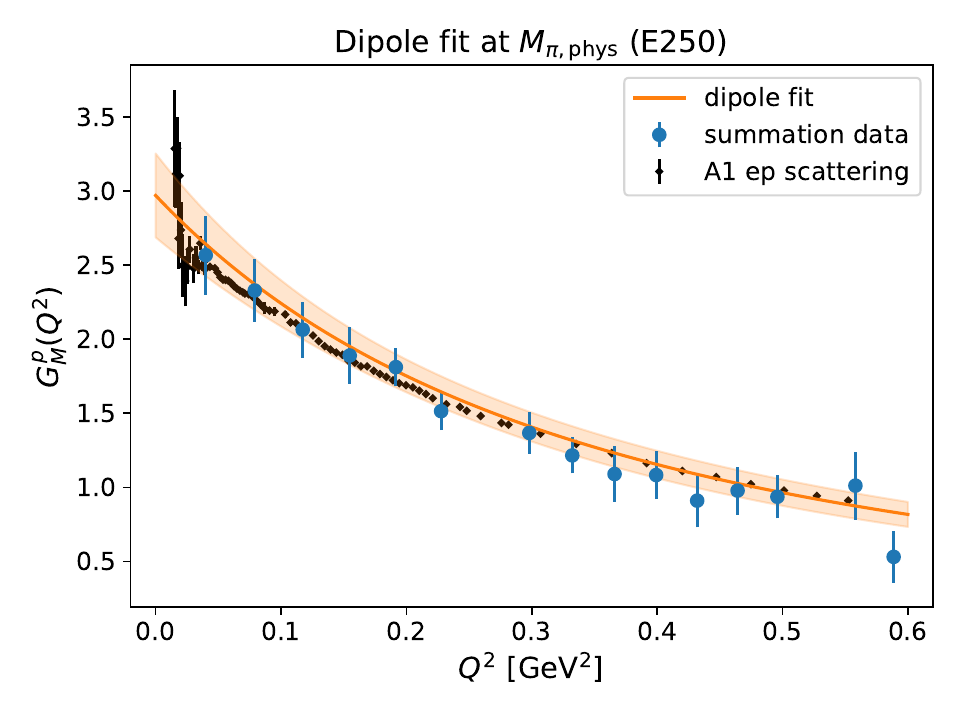}
            \end{center}
        \end{subfigure}
    \end{center}
    \caption{Electromagnetic form factors of the proton as a function of $Q^2$. Our lattice data as obtained from the summation method are represented by the blue points, with the orange curve and band depicting a dipole fit to them and its associated uncertainty. The black diamonds correspond to the experimental $ep$ scattering data from \cite{Bernauer2014} using a classic Rosenbluth separation.}
    \label{fig:ge_gm_proton}
\end{figure}

Having computed the isovector form factors in analogy to \cite{Djukanovic2021} and the isoscalar ones as described in these proceedings, one can combine them to the physical ones for the proton and neutron,
\begin{equation}
    p = \frac{1}{6}(S + 3V) , \quad n = \frac{1}{6}(S - 3V) .
    \label{eq:proton_neutron}
\end{equation}
Since the nucleon is part of the baryon octet, the relevant isoscalar combination is the octet one, $S = u+d-2s$, while $V = u-d$ denotes the isovector combination.
The resulting electromagnetic form factors of the proton at the physical pion mass (E250) are plotted as a function of $Q^2$ in \cref{fig:ge_gm_proton}, together with the respective dipole fits.
Overall, the dipole form seems to describe our data already fairly well.
In the electric form factor, however, there are some minor deviations visible, in particular at the smaller values of $Q^2$.
The figures also include the experimental data from $ep$ scattering \cite{Bernauer2014}.
For $G_M$, these are compatible with our results.
The electric form factor, on the other hand, exhibits a systematic upwards shift of our values compared to the experiment.
This means that our lattice QCD-based estimate of the proton radius will be smaller than that from $ep$ scattering.

\begin{figure}[ht]
    \begin{center}
        \includegraphics[width=0.7\textwidth]{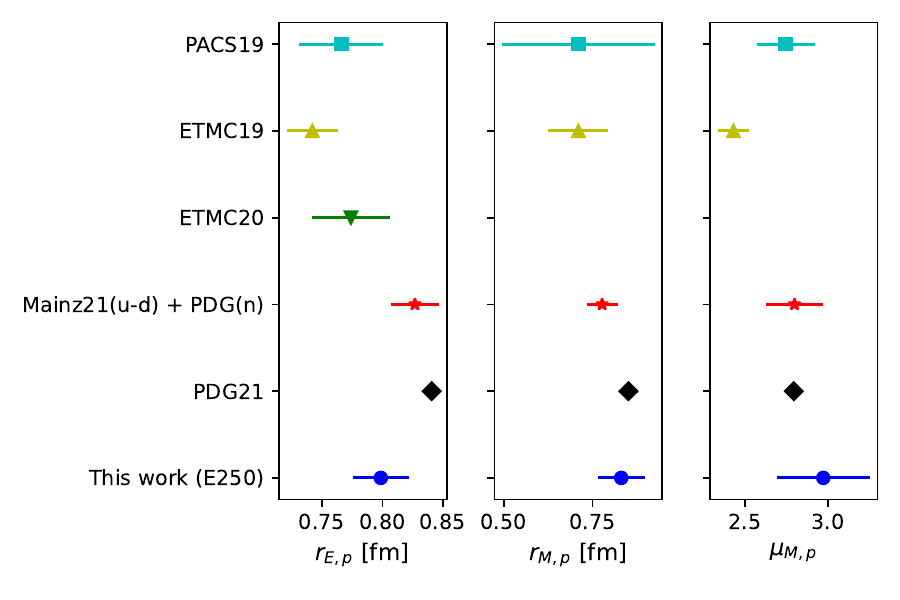}
    \end{center}
    \caption{Comparison of our preliminary results for the electromagnetic charge radii and the magnetic moment of the proton with other studies}
    \label{fig:proton_radii_comparison}
\end{figure}

In \cref{fig:proton_radii_comparison}, we compare our preliminary results for the charge radii and the magnetic moment of the proton, as obtained from a single near-physical ensemble (E250), to a selection of other studies: the 2021 update of the experimental PDG numbers \cite{Zyla2020}, the combination of our isovector results from \cite{Djukanovic2021} with the PDG values for the neutron, along with direct lattice determinations by ETMC \cite{Alexandrou2019,Alexandrou2020} and PACS \cite{Shintani2019}.
For all three quantities, we attain a good agreement between our result and the other lattice investigations as well as the experimental findings.
In the electric radius, there remains a slight tension between our number and the PDG value, despite their 2021 update already favoring a small proton radius.
The point from A1 $ep$ scattering \cite{Bernauer2014} would be even farther to the right here.
Nevertheless, this discrepancy seems to be confirmed by the other lattice determinations shown in \cref{fig:proton_radii_comparison}.
Furthermore, we achieve errors which are comparable with those of similar lattice studies.
Note however that our errors are purely statistical at the moment, whereas for the remaining points in \cref{fig:proton_radii_comparison}, statistical and systematic errors have been added in quadrature, where available.

\section{Conclusions and outlook}
\label{sec:conclusions}
In these proceedings, we have investigated the isoscalar electromagnetic form factors of the nucleon in lattice QCD with $2 + 1$ flavors of dynamical quarks using five different CLS ensembles.
The electromagnetic charge radii and the magnetic moment have been extracted from dipole fits.
Moreover, the proton form factors have been computed on one ensemble with a close-to-physical pion mass.
The corresponding radii and the magnetic moment agree reasonably well with the experimental values and previous lattice determinations, with competitive errors.

In the future, it will be of great interest to extend the analysis presented in \cref{sec:methods} in various directions:
We plan to implement another strategy to deal with the excited-state contamination as an alternative to the \enquote{plain} summation method, allowing for the associated systematic uncertainty to be quantified.
The $Q^2$-dependence of the form factors is commonly parametrized in a model-independent way, for example via a $z$-expansion.
This is expected to alleviate the small yet noticeable deviation of the dipole fits from our data, placing the extracted radii on a firmer ground.
Finally, our results can be extrapolated to the physical point by fitting their $M_\pi^2$- and $a^2$-dependence.

Complementary to these methodological advances, we are working on the expansion of our data set.
There are now smaller source-sink separations available on all the ensembles listed in \cref{tab:ensembles}.
Analyzing further ensembles will give us a clearer picture of the chiral, lattice spacing and finite volume dependence of our observables.
Apart from that, increasing the statistics on the near-physical ensemble E250 and performing additional measurements of the two-point functions for the disconnected contributions would be highly desirable in order to improve on our errors.

\acknowledgments
This research is partly supported by the Deutsche Forschungsgemeinschaft (DFG, German Research Foundation) through project HI 2048/1-2 and through the Cluster of Excellence \enquote{Precision Physics, Fundamental Interactions and Structure of Matter} (PRISMA${}^+$) funded by the DFG within the German Excellence Strategy.
Calculations for this project were partly performed on the HPC clusters \enquote{Clover} and \enquote{HIMster2} at the Helmholtz Institute Mainz, and \enquote{Mogon 2} at Johannes Gutenberg University Mainz (\url{https://hpc.uni-mainz.de}).
The authors also gratefully acknowledge the support of the John von Neumann Institute for Computing and Gauss Centre for Supercomputing e.V. (\url{https://www.gauss-centre.eu}) for projects CHMZ21 and CHMZ36.

\bibliography{literature.bib}
\end{document}